\documentclass[12pt]{article}

\newcommand{\sect}[1]{\setcounter{equation}{0}\section{#1}}

\usepackage{latexsym}
\usepackage{amsfonts}

\def\fraz#1#2{{\strut\displaystyle #1\over\displaystyle #2}}

\def\sca#1#2{({#1}\!\cdot\!{#2})}

\def\chid{\chi_{{}_{D}}}

\def\spazio#1{\vrule height#1em width0em depth#1em}
\def\L{{\cal{L}}}

\date {}
\begin{document}

\title{\bf Foldy-Wouthuysen Transformation for a Spinning Particle with Anomalous Magnetic Moment.}

\author{A. Barducci, R.Giachetti and G. Pettini}

\maketitle \centerline{{  Department of Physics, University of
Florence and I.N.F.N. Sezione di Firenze }}\centerline{{ Via G.
Sansone 1, I-50019 Sesto Fiorentino, Firenze , Italy\footnote{
e-address: barducci@fi.infn.it, giachetti@fi.infn.it, pettini@fi.infn.it} }}
\bigskip
\bigskip

\centerline{{ Firenze Preprint - DFF - 447/01/09}}

\bigskip
\noindent

\begin{abstract}
In this paper we study the Foldy Wouthuysen transformation for
a pseudoclassical particle with anomalous magnetic moment in an
external, stationary electromagnetic field. We show that the transformation
can be expressed in a closed form for neutral particles in 
purely electrostatic fields and for neutral and charged particles in external magnetostatic
fields. The explicit expressions of the diagonalized Hamiltonians are calculated.
\end{abstract}

\bigskip

\sect{Introduction.}

Pseudoclassical spinning relativistic  particles and superparticles have found in past years different descriptions
related both  to electromagnetic interactions,  \cite{BM}-\cite{GT},  and to interactions of a more general nature \cite{BCL2,BCL3}.
These models, firstly connected with properties emerging from
string theory, later on became interesting in themselves and their quantum structure was thoroughly
investigated, for example by determining the Foldy-Wouthuysen transformation (hereafter FWT) in the presence of a stationary
magnetic field. \cite{BCL4}.
We recall that the FWT is based on a canonical transformation which reduces the wave equation to a representation 
where the Hamiltonian is an even matrix, usually in the form of a square-root operator containing both kinetic and interaction energy.
 More refined results were obtained when the external fields were taken as plane waves.
In this case, in analogy with well established classical results \cite{VL}, it was proven by path integral 
\cite{BG75}-\cite{BGJP} and by canonical theory \cite{BGJP05,BGJP2}
that the semi-classical approximation reproduced the exact quantum propagator.
The description of spinning particles was also generalized by allowing for the presence of an anomalous magnetic moment,
firstly introduced  in \cite{B} and subsequently in \cite{GS1}-\cite{BBC2001}, all of these
treatments leading to the same first class Dirac constraints and hence to the same wave equation.
More recently, by using this approach, we considered the quantization of a spinning particle with 
anomalous magnetic moment in the field of an  electromagnetic plane wave \cite{BGJP2}, generalizing
the results  obtained in \cite{BGJP}.  We found that the semi-classical approximation is no longer exact, but for
some particular cases, due to the effects arising from the interference of the anomalous magnetic moment with the electric charge that requires
the presence of a $T$-ordered product in the quantum propagator.

The purpose of this work is to extend the FWT to a
pseudoclassical spinning particle with  anomalous magnetic moment in a
stationary electromagnetic field, thus completing the research begun in \cite{BCL4} for the usual spinning particle. 
We are particularly  interested in studying the cases in which the result can be expressed in a closed form. These
turn out to be the following: $(a)$ a neutral particle in a stationary electric field; $(b)$ a neutral particle in a 
stationary magnetic field; $(c)$ a charged particle in a stationary magnetic field.  The technique that yields the
results is not so different from the one we used, for example,  in \cite{BGJP2}. A crucial point that
distinguishes this paper from our previous ones, however, is the need to exploit two possible different
ways of realizing the quantization of the Dirac pseudoclassical brackets:
in fact these representations of the Clifford algebra, coming from the quantization of the pseudoclassical variables, 
give rise to two different expressions for the Dirac equation, intertwined by a Pauli-Gursey unitary transformation,  \cite{BBC,PG},  that allow us to
treat more leisurely the different interacting cases.

The content of this paper can be summarized as follows.  In Section \ref{FWT-General} we briefly recall the 
quantization scheme of  the pseudoclassical particle with anomalous magnetic moment: we write down the singular
Lagrangian, the Dirac constraints and their corresponding operator form leading to the wave equation. 
We then formulate the FWT problem  and we
consider its general features. In Section \ref{FWT-Results} we present a detailed discussion of the results concerning the above mentioned three cases
in which the transformation can be expressed in a closed form.

\bigskip
\sect{The general setting of the FWT for the spinning particle with anomalous magnetic moment.}
\label{FWT-General}

For the sake of completeness in this Section we briefly summarize our notations and the Dirac constraints leading to the
canonical quantization of the pseudoclassical particle with anomalous magnetic moment. The details can be found in
\cite{BGJP2,B}. With the usual conventions for the metric tensor and for the
gamma matrices \cite{bjorkendrell}, in a unit system with $\hbar=c=1$, the Lagrangian we start with is \cite{B}
\begin{eqnarray}
&{}& \L(x_\mu,\dot x_\mu,\xi_\mu,{\dot \xi}_\mu,\xi_5,{\dot
\xi}_5)=\spazio{1.0}\cr &{}&-\fraz i2 \sca\xi{\dot \xi}-\fraz i2
\xi_5{\dot \xi}_5-q\,\sca{\dot x} A - \Bigl[{m^2
-i(q+\frac{e\mu}2)\,F_{\mu\nu}\xi^\mu\xi^\nu
-\frac{e^2\mu^2}{16m^2}\,F_{\mu\nu}F_{\rho\sigma}\,\xi^\mu\xi^\nu\xi^\rho\xi^\sigma}\Bigr]^{1/2}\spazio{1.0}\cr
&{}& \phantom{XXXXXX} \Bigl[{\Bigl({\dot
x}^\mu-i\,(m+\frac{ie\mu}{4m}\,F_{\lambda\nu}\,\xi^\lambda\xi^\nu)^{-1}
\,\,\xi^\mu\,({\dot\xi}_5-\frac{e\mu}{2m}\,{\dot
x}^\rho\,F_{\rho\sigma}\,\xi^\sigma)\Bigr)^2\,
}\Bigr]^{1/2} \label{spinlagrangian}
\end{eqnarray}
where  $\mu$ is related to the anomalous magnetic moment, $q$ is the charge of the particle and $e$ the electronic charge respectively. The Lagrangian (\ref{spinlagrangian}) is
evidently singular and gives rise to the two first class constraints
\begin{eqnarray}
&{}&\chid=\sca{\Pi}\xi-m\, \xi_5+i\frac{e\mu}{4m}\,F_{\mu\nu}\,\xi^\mu\xi^\nu\xi_5\,,
\spazio{1.2}\cr
\nonumber
\label{constr2b}
&{}&\chi=\Pi^2-m^2+i(q+\frac {e\mu}2) F_{\mu\nu}\,\xi^\mu\xi^\nu
+i\frac{e\mu}{m}\,\Pi^\mu\,F_{\mu\nu}\,\xi^\nu\xi_5+
\frac{e^2\mu^2}{16m^2}\,F_{\mu\nu}F_{\rho\sigma}\,\xi^\mu\xi^\nu\xi^\rho\xi^\sigma,
\nonumber
\label{constr1b}
\end{eqnarray}
where the kinetic momentum $\Pi$ is related to the canonical momentum $p$  by 
\begin{eqnarray}
\Pi^\mu=p^\mu-qA^\mu 
\label{Pi}
\end{eqnarray}
and the second class constraints have already been accounted for. Their algebra
\begin{eqnarray}
~~~~~~~~~~\!\!\!\!\!\!\!\!\!\!\!\!\!\!\!\!\!\!\{\chid,\chid\}=i\chi,\qquad\{\chid,\chi\}=0,\qquad\{\chi,\chi\}=0 \nonumber
\label{constralg1}
\end{eqnarray}
is determined by the nonvanishing Dirac brackets of the pseudoclassical variables,
\begin{eqnarray}
~~~~~~~~~~\!\!\!\!\{x^\mu,p^\nu\}=-\eta^{\mu\nu}\qquad~~\!\{\xi^\mu,\xi^\nu\}=i\eta^{\mu\nu},\qquad~~\{\xi_5,\xi_5\}=-i,
\nonumber
\label{DirBra}
\end{eqnarray}
that, upon quantization, give rise to the graded commutators
\begin{eqnarray}
~~~~~~~~~~[x^\mu,p^\nu]=-i\eta^{\mu\nu},\qquad
\{{\hat \xi}^\mu,{\hat \xi}^\nu\}_+=-\eta^{\mu\nu},\!\qquad\{{\hat \xi}_5,{\hat \xi}_5\}_+=1
\label{constralg2}
\end{eqnarray}

It was previously observed \cite{BCL1,BBC,BGJP} that the anti-commutation relations (\ref{constralg2}) of the odd operators $\hat \xi^\mu$, $\hat\xi_5$, can be satisfied by two different representations 
\begin{eqnarray}
&{}& {\hat \xi}^\mu=2^{-1/2}\,\gamma_5\gamma^{\mu},\phantom{i}\qquad\qquad {\hat \xi}_5=2^{-1/2}\,\gamma_5\spazio{1.0}
\label{xigamma1}\\
&{}& {\hat \xi}^\mu=2^{-1/2}\,i\,\gamma^{\mu},\phantom{\gamma_5}\qquad\qquad {\hat \xi}_5=2^{-1/2}\,\gamma_5
\label{xigamma2}
\end{eqnarray}
It was also observed, \cite{BBC}, that the two representations are connected by a Pauli-Gursey transformation,\textit{ i.e.} a conjugation by the
matrix $\exp[ i(\pi/4)\gamma_5]$.
However, contrary to almost all the previously quoted papers, where only  the representation (\ref{xigamma1}) was effectively used, in the following both realizations will appear, since the different cases we will examine are treated more efficiently if the appropriate choice is made. The
 explicit form of the quantized Dirac Hamiltonian in the realization   (\ref{xigamma1}) takes
the form
\begin{eqnarray}
\hat{H}_{{}_D}=({\overrightarrow{\alpha}}\!\cdot\!\overrightarrow{\Pi})+qA_0+\beta m+\frac{e\mu}{8m}\beta\sigma_{\mu\nu}F^{\mu\nu}
\label{constralgD}
\end{eqnarray}
where $\beta=\gamma^0\,,~\overrightarrow{\alpha}=\gamma^0\overrightarrow{\gamma}$ and
$\sigma_{\mu\nu}=(i/2)[\gamma_\mu,\gamma_\nu]$, \cite{bjorkendrell}.  In the realization   (\ref{xigamma2}) we have, instead,
\begin{eqnarray}
\hat{H}_{{}_{PG}}= (\overrightarrow{\alpha}\!\cdot\!\overrightarrow{\Pi})+qA_0-i\beta\gamma_5 m-\frac{ie\mu}{8m}\beta\gamma_5\sigma_{\mu\nu}F^{\mu\nu}
\label{constralgPG}
\end{eqnarray}
We can easily verify that
\begin{eqnarray}
e^{ i\pi\gamma_5/4}\,\hat{H}_{{}_D}\,e^{-i\pi\gamma_5/4}=\hat{H}_{{}_{PG}}\nonumber
\end{eqnarray}
In view of the discussion of the FWT, we find it useful to separate, both in $\hat{H}_{{}_D}$ and in $\hat{H}_{{}_{PG}}$, the even and the odd terms. We therefore write
\begin{eqnarray}
\hat{H}_{{}_D}=\hat{H}_{{}_D}^{{even}}+\hat{H}_{{}_D}^{{odd}}\,,\qquad\qquad \hat{H}_{{}_{PG}}=\hat{H}_{{}_{PG}}^{{even}}+\hat{H}_{{}_{PG}}^{{odd}}\,,
\nonumber
\end{eqnarray}
and making explicit the electromagnetic tensor $F^{\mu\nu}$ in (\ref{constralgD}) and (\ref{constralgPG}) we obtain
\begin{eqnarray}
&{}\!\!\!\!\!\!\!\!\!\!\!\!\! \hat{H}_{{}_D}^{{even}}=qA_0+\beta m-{\displaystyle\frac{e\mu}{4m}}\beta\,(\overrightarrow{\Sigma}\!\cdot\!\overrightarrow{B}) \qquad~~~~~~~\qquad
\hat{H}_{{}_D}^{{odd}}=(\overrightarrow{\alpha}\!\cdot\!\overrightarrow{\Pi})+{\displaystyle\frac{ie\mu}{4m}}\beta\,(\overrightarrow{\alpha}\!\cdot\!\overrightarrow{E})
\label{HDeven-odd}\\
&{}\!\!\!\!\!\!\!\!\!\!\!\!\! \hat{H}_{{}_{PG}}^{{even}}=qA_0+{\displaystyle \frac{e\mu}{4m}}\beta\,\gamma_5(\overrightarrow{\alpha}\!\cdot\!\overrightarrow{E})\qquad\quad
\hat{H}_{{}_{PG}}^{{odd}}=(\overrightarrow{\alpha}\!\cdot\!\overrightarrow{\Pi})-im\beta\,\gamma_5+{\displaystyle\frac{ie\mu}{4m}}\beta\,\gamma_5\,(\overrightarrow{\Sigma}\!\cdot\!\overrightarrow{B})
\label{HPGeven-odd}
\end{eqnarray}
where the spatial spin vector $\overrightarrow\Sigma$, defined by the relation $\sigma^{ij}=\epsilon^{ijk}\Sigma^k$, can also be written as $\overrightarrow\Sigma=\gamma_5\overrightarrow\alpha$.

It is well known that in the simplest systems, as for example the Dirac free particle and the Dirac particle  with no anomalous magnetic moment
in a magnetic field, the FWT depends upon the odd part of the Hamiltonian that, in the two mentioned cases, is given by the kinetic part $(\overrightarrow{\alpha}\!\cdot\!\overrightarrow{\Pi})$ and anti-commutes
with the even part $\beta m$.  We will follow a similar method also for the more general interacting case with the additional difficulties which we will discuss later on. We therefore give an extremely rapid summary of the successive steps necessary to get the result in these two cases, starting from the non-interacting case, obtained by (\ref{constralgD}-\ref{constralgPG}) with $\mu=e=0$. Since the FWT is generally determined by  $\exp\bigl[\beta{\cal O}\bigr]$,  ${\cal O}$ being the odd part of the Hamiltonian operator, \cite{bjorkendrell,FW},
it can be seen that for the first realization  (\ref{xigamma1})  of the Clifford algebra, the unitary transformation  is generated by
\begin{eqnarray}
 \exp\bigl[i\hat S_{{}_D}\bigr]=\exp\bigl[ \beta (\overrightarrow{\alpha}\!\cdot\!\overrightarrow{p}) \theta(|\overrightarrow{p}|)\bigr]\qquad {\mathrm {where}}\qquad
\theta(|\overrightarrow{p}|)=\frac 1{2|\overrightarrow{p}|}\arctan \frac{|\overrightarrow{p}|}m
\label{FWT_D}
\end{eqnarray}
The FW transformed Hamiltonian operator is then
\begin{eqnarray}
{{\widetilde {\hat H}}_{{}_D}}=e^{ \displaystyle \beta (\overrightarrow{\alpha}\!\cdot\!\overrightarrow{p}) \theta(|\overrightarrow{p}|)}\,\Bigl((\overrightarrow{\alpha}\!\cdot\!\overrightarrow{p})+\beta m\Bigr)\,e^{ \displaystyle -\beta (\overrightarrow{\alpha}\!\cdot\!\overrightarrow{p}) \theta(|\overrightarrow{p}|)}=\beta\Bigl[{\overrightarrow{p}}^2+m^2\Bigr]^{1/2}
\nonumber
\end{eqnarray}
For the  realization (\ref{xigamma2})  of the Clifford algebra the whole Hamiltonian $\hat{H}_{{}_{PG}}$ is odd, as the mass term $-im\beta\gamma_5$ is itself odd (and still anti-commuting with $(\overrightarrow\alpha\!\cdot\!\overrightarrow\Pi)$). The expression (\ref{FWT_D}) has therefore to be substituted by
\begin{eqnarray}
 \exp\bigl[i\hat S_{{}_{PG}}\bigr]=\exp\Bigl[ \beta \Bigl((\overrightarrow{\alpha}\!\cdot\!\overrightarrow{p}) -im\beta\gamma_5\Bigr)\phi(|\overrightarrow{p}|)\Bigr]\quad {\mathrm {where}}\quad
\phi(|\overrightarrow{p}|)=\frac {\pi}{4\,\bigl[{\overrightarrow{p}}^2+m^2 \bigr]^{1/2}}
\nonumber
\end{eqnarray}
As expected, the transformed Hamiltonian reads again ${{\widetilde {\hat H}}_{{}_{PG}}}=\beta\bigl[{\overrightarrow{p}}^2+m^2\bigr]^{1/2}\,$.

The Hamiltonian operators for the pseudoclassical particle interacting with  a stationary magnetic field in the two representations are obtained from (\ref{HDeven-odd}$\,$-$\,$\ref{HPGeven-odd}) by choosing $\mu=0$,  $A_0=0$
and $\overrightarrow A=\overrightarrow A(\overrightarrow x)$. The computations are somewhat more cumbersome, but can still be managed and
give a result in a closed form. We introduce
\begin{eqnarray}
\hat{\Lambda}=-\{(\hat{ \overrightarrow \xi}\!\cdot\!\overrightarrow \Pi),(\hat{ \overrightarrow \xi}\!\cdot\!\overrightarrow \Pi)\}_+
=\frac 12\{({ \overrightarrow \gamma}\!\cdot\!\overrightarrow \Pi),({ \overrightarrow \gamma}\!\cdot\!\overrightarrow \Pi)\}_+=
-\Bigl({\overrightarrow \Pi}^2+\frac q2
\sigma^{ij}F^{ij}\Bigr)
\nonumber
\end{eqnarray}
and by quantizing the graded Jacobi identity
\begin{eqnarray}
 \sum\limits_{\mathrm{cyclic}} (-1)^{d_\ell\,d_n}\,\{v_\ell,\{v_m,v_n\}\}=0
\nonumber
\end{eqnarray} 
where $v_i$ is a generic dynamical variable of degree $d_i\!=\!0,1$ according to its parity,
we easily verify that $[\beta(\overrightarrow \alpha\!\cdot\!\overrightarrow \Pi),\hat{\Lambda}]=0$. By a direct calculation it can then be proved that the similarity transformation of ${\hat H}_{{}_D}$ with
\begin{eqnarray}
\exp\bigl[i\hat S_{{}_D}\bigr]=\exp\Bigl[ \beta (\overrightarrow{\alpha}\!\cdot\!\overrightarrow{\Pi})\, \theta(\hat{\Lambda})\Bigr]\qquad {\mathrm {where}}\qquad
\theta(\hat{\Lambda})=\frac 1{2\sqrt{\hat{\Lambda}}}\arctan \frac{\sqrt{\hat{\Lambda}}}m 
\nonumber
\end{eqnarray}
gives the Hamiltonian \cite{BCL4}
\begin{eqnarray}
{{\widetilde {\hat H}}}_{{}_D}=\beta\Bigl[{\overrightarrow\Pi}^2-q(\overrightarrow{\Sigma}\!\cdot\!\overrightarrow{B})+m^2\Bigr]^{1/2}
\label{H_mag}
\end{eqnarray}
Note that (\ref{H_mag}) is the same expression obtained for a stationary and uniform magnetic field, as found in \cite{TW}.
The previous result can be also deduced  by transforming (\ref{HPGeven-odd}), where now the complete Hamiltonian is odd, by
\begin{eqnarray}
\exp\bigl[i\hat S_{{}_{PG}}\bigr]=\exp\Bigl[ \beta \Bigl((\overrightarrow{\alpha}\!\cdot\!\overrightarrow{\Pi})-im\beta\gamma_5 \Bigr)\phi(\hat{\Lambda})\Bigr]\qquad {\mathrm {where}}\qquad
\phi(\hat{\Lambda})=\frac {\pi}4 \,\frac1{\sqrt{-\hat{\Lambda}+m^2} }
\nonumber
\end{eqnarray}
and in this case the algebra is simpler as the whole Hamiltonian is odd, so that $ {\mathcal O}=\hat{H}_{{}_{PG}}$.

The difficulties that arise in the general interacting case, both for $\hat{H}_{{}_{D}}$ and $\hat{H}_{{}_{PG}}$, are that the kinetic part does not anti-commute with the even terms $qA_0$, $\beta\,(\overrightarrow\Sigma\!\cdot\!\overrightarrow{B})$, $\beta\,\gamma_5(\overrightarrow{\alpha}\!\cdot\!\overrightarrow{E})$ and that the interaction parts contain terms of both even and odd parity.
Starting with these premises, in the next section we are going to examine the cases in which explicit results can be reached.

\bigskip
\sect{The discussion of the results.}
\label{FWT-Results}

We begin our report on the cases admitting a complete and closed solution by examining a neutral particle, $q=0$. The Hamiltonian operators in the two different representations are given by (\ref{HDeven-odd}-\ref{HPGeven-odd}) and both of them involve even and odd terms in the interaction part. We therefore proceed by separating the electrostatic from the magnetostatic interaction.
\medskip

$(a)$ $q=0$,\textit{ pure electrostatic interaction}. Consider first the case with $\overrightarrow{B}=0$. We have
\begin{eqnarray}
\hat{H}_{{}_{D}}={\mathcal O}+{\mathcal E} ,\qquad {\mathcal O}=(\overrightarrow{\alpha}\!\cdot\!\overrightarrow{P}),
 \qquad {\mathcal E}=\beta m, \qquad {\mathrm{where}}  \quad 
\overrightarrow{P}= \overrightarrow{p}-\frac{ie\mu}{4m}\beta\overrightarrow{E}
\label{q0B0}
\end{eqnarray}
In (\ref{q0B0}) we have denoted by ${\mathcal O},\,{\mathcal E}$ the odd and the even term respectively. We then see that the structure is extremely similar to that of the free particle, but for the translation of the momentum by a factor linear in the electric field. Moreover, since
\begin{eqnarray}
\bigl[\,P_i  ,P_j   \bigr] = -\frac{e\mu}{4m}\beta\Bigl( \frac{\partial E_j}{\partial x_i}- \frac{\partial E_i}{\partial x_j}\Bigr)
\nonumber
\end{eqnarray}
for a conservative field, $\overrightarrow{\nabla}\times\overrightarrow{E}=0$, our procedure could be considered a canonical transformation
with respect to the free case.
In the general case, the FWT will be generated by an exponential $\exp\{\beta{\mathcal O}\varphi\}$, where $\varphi$ is a parameter to be determined
in order to obtain a totally even transformed Hamiltonian. Observe now that a straightforward calculation gives the form of the even term 
\begin{eqnarray}
 (\beta{\mathcal O})^2=-{\mathcal O}^2=-\Bigl[\overrightarrow{p}^2+\Bigl(\frac{e\mu}{4m}\overrightarrow{E}\Bigr)^2-
\frac{e\mu}{4m}\beta\Bigl(\overrightarrow{\nabla}\!\cdot\!\overrightarrow{E}+\overrightarrow{\Sigma}\!\cdot\!(\overrightarrow{E}\!\times\!\overrightarrow{p}-
\overrightarrow{p}\!\times\!\overrightarrow{E})
\Bigr)\Bigr]
\label{O2_ES}
\end{eqnarray}
Moreover, by parity properties,
\begin{eqnarray}
 \bigl[\,\beta{\mathcal O},\beta\,\bigr]_+=\bigl[\,\beta{\mathcal O},{\mathcal O}\,\bigr]_+=0, \qquad
\bigl[\,\beta{\mathcal O},{\mathcal O}^2\,\bigr]_-=0
\label{oddAC}
\end{eqnarray}
it is easily seen that
\begin{eqnarray}
 (\beta{\mathcal O})^3=-\beta{\mathcal O}^3,\quad (\beta{\mathcal O})^4={\mathcal O}^4,\quad (\beta{\mathcal O})^5=\beta{\mathcal O}^5
\quad {\mathrm{and}}~ {\mathrm{so}}~ {\mathrm{on}}.
\label{powerodd}
\end{eqnarray}
We therefore find
\begin{eqnarray}
{{\widetilde {\hat H}}_{{}_D}}=e^{ \displaystyle \beta {\mathcal O}\varphi}\, \Bigl[(\overrightarrow{\alpha}\!\cdot\!\overrightarrow{\Pi})
+\beta m\Bigr] \,e^{ -\displaystyle \beta {\mathcal O}\varphi}=\Bigl[(\overrightarrow{\alpha}\!\cdot\!\overrightarrow{\Pi})
+\beta m\Bigr] \,e^{ -\displaystyle 2 \beta {\mathcal O}\varphi}
\nonumber
\label{FWES}
\end{eqnarray}
Due to (\ref{oddAC}$\,$-$\,$\ref{powerodd}) the exponential is easily calculated and yields
\begin{eqnarray}
 e^{ -\displaystyle 2 \beta {\mathcal O}\varphi}=\cos(2\sqrt{{\mathcal O}^2}\varphi)-\frac{\beta{\mathcal O}}{\sqrt{{\mathcal O}^2}}\sin(2\sqrt{{\mathcal O}^2}\varphi)
\label{e2bO}
\end{eqnarray}
so that, expanding the expression of the transformed Hamiltonian, we find
\begin{eqnarray}
{{\widetilde {\hat H}}_{{}_D}}=  {\mathcal O}\Bigl[\cos(2\sqrt{{\mathcal O}^2}\varphi)-\frac{m}{\sqrt{{\mathcal O}^2}}\sin(2\sqrt{{\mathcal O}^2}\varphi)\Bigr]
+\beta\Bigl[{\sqrt{{\mathcal O}^2}}\sin(2\sqrt{{\mathcal O}^2}\varphi)+m\cos(2\sqrt{{\mathcal O}^2}\varphi)\Bigr]
\nonumber
\end{eqnarray}
Finally, by choosing
\begin{eqnarray}
 \varphi=\frac{1}{2\sqrt{{\mathcal O}^2}}\,\arctan\Bigl( \frac{\sqrt{{\mathcal O}^2}}{m}\Bigr)
\nonumber
\end{eqnarray}
we find a completely even transformed Hamiltonian, whose final form is
\begin{eqnarray}
  \!\!\!\!\!{{\widetilde {\hat H}}_{{}_D}}\!=\!\beta\Bigl[ {\mathcal O}^2+m^2  \Bigr]^{1/2}
\!\!=\!
\beta\Bigl[ 
\overrightarrow{p}^2\!+\!\Bigl({\displaystyle\frac{e\mu}{4m}}  \overrightarrow{E} \Bigr)^2\!\!-\!{\displaystyle\frac{e\mu}{4m}}\beta\Bigl(
(\overrightarrow{\nabla}\!\cdot\!\overrightarrow{E})\!+\!(\overrightarrow{\Sigma}\!\cdot\!(\overrightarrow{E}\!\times\!\overrightarrow{p}\!-\!\overrightarrow{p} \!\times\!\overrightarrow{E}
))
 \Bigr)
\Bigr]^{1/2}
\label{HDq0B0}
\end{eqnarray}
\medskip

$(b)$ $q=0$,\textit{ pure magnetostatic interaction}. The other case we can discuss with $q=0$ is the pure magnetostatic interaction. The Dirac Hamiltonian is
\begin{eqnarray}
{{ {\hat H}}_{{}_D}}\!=\! (\overrightarrow{\alpha}\!\cdot\!\overrightarrow{p})+\beta m-\frac{e\mu}{4m}\beta (\overrightarrow{\Sigma}\!\cdot\!\overrightarrow{B})
\nonumber
\end{eqnarray}
and in it the interaction term is even. If, however, we consider the same problem in the Pauli-Gursey representation, we have a completely odd
Hamiltonian
\begin{eqnarray}
{{ {\hat H}}_{{}_{PG}}}\!=\!  (\overrightarrow{\alpha}\!\cdot\!\overrightarrow{p})-im\beta\gamma_5+\frac{ie\mu}{4m}\beta  (\overrightarrow{\alpha}\!\cdot\!\overrightarrow{B})
\label{HPGq0E0}
\end{eqnarray}
Since the choice of a particular representation is irrelevant with respect to the FWT,  it is certainly more convenient to start with 
${\mathcal O}={{ {\hat H}}_{{}_{PG}}}$ given in (\ref{HPGq0E0}). As usual we will consider a similarity transformation
generated by $\exp(\beta{\mathcal O}\varphi)$, looking, as we previously did, whether we can also satisfy the further requirements 
$[\beta,\varphi]_-=[{\mathcal O},\varphi]_-=0$: if this is the case, we will be able to give a closed form to the FWT and to the transformed Hamiltonian as we did in the previous paragraph. We will give an \textit{a posteriori} solution to these questions.

In analogy with (\ref{O2_ES}) we first calculate the even term
\begin{eqnarray}
 (\beta{\mathcal O})^2=-{\mathcal O}^2\!=\!-\Bigl[\overrightarrow{p}^2\!+\!m^2\!+\!\Bigl(\frac{e\mu}{4m}\overrightarrow{B}\Bigr)^2\!-\!\frac{e\mu}{2}
(\overrightarrow{\Sigma}\!\cdot\!\overrightarrow{B})
\!-\!\frac{e\mu}{4m}\beta\Bigl(\overrightarrow{\Sigma}\!\cdot\!(\overrightarrow{B}\!\times\!\overrightarrow{p}\!-\!
\overrightarrow{p}\!\times\!\overrightarrow{B})
\Bigr)\Bigr]
\label{O2_MS}
\end{eqnarray}
and the relations (\ref{oddAC}) hold in this case too. Therefore
\begin{eqnarray}
{{\widetilde {\hat H}}_{{}_{PG}}}=e^{ \displaystyle \beta {\mathcal O}\varphi}\,{ {\hat H}}_{{}_{PG}} \,e^{ -\displaystyle \beta {\mathcal O}\varphi}= \Bigl((\overrightarrow{\alpha}\!\cdot\!\overrightarrow{p})-im\beta\gamma_5+\frac{ie\mu}{4m}\beta  (\overrightarrow{\alpha}\!\cdot\!\overrightarrow{B})\Bigr) \,e^{ -\displaystyle 2 \beta {\mathcal O}\varphi}
\nonumber
\label{FWMS}
\end{eqnarray}
and since  $\exp({ -\displaystyle 2 \beta {\mathcal O}\varphi})$ is again given by (\ref{e2bO}), if we choose 
\begin{eqnarray}
\varphi=\frac\pi 4\,\frac1{\sqrt{{\mathcal O}^2}}
\label{Pi/4}
\end{eqnarray}
we find an explicit form for the FW transformed Hamiltonian, that results in
\begin{eqnarray}
   \!\!\!\!\!{{\widetilde {\hat H}}_{{}_{PG}}}\!=\!\beta\sqrt{ {\mathcal O}^2}
\!\!=\!\beta
\Bigl[ 
\overrightarrow{p}^2\!+\!m^2\!+\!\Bigl({\displaystyle\frac{e\mu}{4m}}  \overrightarrow{B} \Bigr)^2\!\!-\!\frac{e\mu}{2}
(\overrightarrow{\Sigma}\!\cdot\!\overrightarrow{B})
\!-\!\frac{e\mu}{4m}\beta\Bigl(\overrightarrow{\Sigma}\!\cdot\!(\overrightarrow{B}\!\times\!\overrightarrow{p}\!-\!
\overrightarrow{p}\!\times\!\overrightarrow{B})
\Bigr)
\Bigr]^{1/2}
\label{HDq0E0}
\end{eqnarray}
One can finally verify that all of our working hypotheses are satisfied.
\medskip

$(c)$ $q\not=0$\textit{ pure magnetostatic interaction}. Let us now turn to the case $q\not=0$. A closed form for the FW transformed Hamiltonian can be found only when $A_0=0$. Since, moreover,
we are considering a stationary case, $\partial\overrightarrow{A}(t)/\partial t=0$, our assumption corresponds to a vanishing electric field.
This is very reasonable from a physical point of view, as a non-vanishing $\overrightarrow{E}$ could lead to the pair production phenomenon
and therefore to the mixing of `large' and `small' components: indeed it is well known that when $\overrightarrow{E}\not=0$, even for a
vanishing magnetic field and an anomalous magnetic moment, the FWT cannot be put in a closed form. The model we are now discussing can thus 
describe a proton in a magnetostatic field. We report here the two representations of the Hamiltonians of the system in terms of $\gamma$-matrices:
\begin{eqnarray}
 &{}& {{ {\hat H}}_{{}_{D}}}\!=\!\beta\Bigl(\overrightarrow{\gamma}\!\cdot\!(\overrightarrow{p}-q\overrightarrow{A})+m\Bigr)
-\frac{e\mu}{4m}\beta (\overrightarrow{\Sigma}\!\cdot\!\overrightarrow{B})
\spazio{1.2}\cr
&{}& {{ {\hat H}}_{{}_{PG}}}\!=\!\beta\,\,\overrightarrow{\gamma}\!\cdot\!(\overrightarrow{p}-q\overrightarrow{A})-im\beta\gamma_5+
\frac{ie\mu}{4m}(\overrightarrow{\gamma}\!\cdot\!\overrightarrow{B})
\label{Hqnot0}
\end{eqnarray}
As in item $(b)$ the second relation in (\ref{Hqnot0}) shows that the Hamiltonian in the Pauli-Gursey representation is
completely odd (\textit{ i.e.} $ {{ {\hat H}}_{{}_{PG}}}\!=\!{\mathcal O}$) and will be more conveniently used for the FWT. Here again we can establish a relation similar to
(\ref{O2_ES}) and (\ref{O2_MS}), that reads
\begin{eqnarray}
 (\beta{\mathcal O})^2=-{\mathcal O}^2\!=\!-\Bigl[\overrightarrow{\Pi}^2\!+\!m^2\!+\!\Bigl(\frac{e\mu}{4m}\overrightarrow{B}\Bigr)^2\!-\!\Bigl(q\!+\!\frac{e\mu}{2}\Bigr)
(\overrightarrow{\Sigma}\!\cdot\!\overrightarrow{B})
\!-\!\frac{e\mu}{4m}\beta\Bigl(\overrightarrow{\Sigma}\!\cdot\!(\overrightarrow{B}\!\times\!\overrightarrow{\Pi}\!-\!
\overrightarrow{\Pi}\!\times\!\overrightarrow{B})
\Bigr)\Bigr]
\label{O2_gen}
\nonumber
\end{eqnarray}
where $\Pi^\mu$ is the canonical momentum (\ref{Pi}). The relations  (\ref{oddAC}) and (\ref{powerodd}) hold in this case too, so that we can directly write
\begin{eqnarray}
{{\widetilde {\hat H}}_{{}_{PG}}}\!=\! {\hat H}_{{}_{PG}}\,e^{ -\displaystyle 2 \beta {\mathcal O}\varphi}
\end{eqnarray}
and with the choice (\ref{Pi/4}) for the angle $\varphi$ we get the final form of the transformed Hamiltonian
\begin{eqnarray} 
 \!{{\widetilde {\hat H}}_{{}_{PG}}}\!=\!\beta\sqrt{ {\mathcal O}^2}
\!\!=\!\beta
\Bigl[ 
\overrightarrow{\Pi}^2\!+\!m^2\!+\!\Bigl(\frac{e\mu}{4m}\overrightarrow{B}\Bigr)^2\!\!-\!\Bigl(q\!+\!\frac{e\mu}{2}\Bigr)
(\overrightarrow{\Sigma}\!\cdot\!\overrightarrow{B})
\!-\!\frac{e\mu}{4m}\beta\Bigl(\overrightarrow{\Sigma}\!\cdot\!(\overrightarrow{B}\!\times\!\overrightarrow{\Pi}\!-\!
\overrightarrow{\Pi}\!\times\!\overrightarrow{B})
\Bigr)
\Bigr]^{1/2}\!\!\cr
\label{HDq_gen}
\end{eqnarray}

We can conclude that the analysis of the pseudoclassical mechanics is quite useful in the derivation of new results. In fact in the present case,
starting from the pseudoclassical mechanics and analyzing the possible different representations of the Clifford algebra,  arising from the quantization of the pseudoclassical Grassmann variables,  we have shown how to extend in a very simple way the quantum unitary transformation which diagonalizes the Dirac Hamiltonian for a particle with anomalous magnetic moment interacting with a stationary non-homogeneous electromagnetic field.

\bigskip
\bigskip

\vfill\break


\end{document}